\documentclass{PoS}

\title{Prospects of the SHiP and NA62 experiments at CERN for hidden sector searches}

\ShortTitle{Hidden sector searches with SHiP and NA62}

\author{\speaker{Philippe Mermod}, on behalf of the SHiP Collaboration \\
        Particle Physics Department, Faculty of Science, University of Geneva, Geneva, Switzerland \\
        E-mail: \email{philippe.mermod@cern.ch}}

\abstract{High-intensity proton beams impinging on a fixed target or beam dump allow to probe new physics via the production of new weakly-coupled particles in hadron decays. The CERN SPS provides opportunities to do so with the running NA62 experiment and the planned SHiP experiment. Reconstruction of kaon decay kinematics (beam mode) allows NA62 to probe for the existence of right-handed neutrinos and dark photons with masses below 0.45~GeV. Direct reconstruction of displaced vertices from the decays of new neutral particles (dump mode) will allow NA62 and SHiP to probe right-handed neutrinos with masses up to 5~GeV and mixings down to several orders of magnitude smaller than current constraints, in regions favoured in models which explain at once neutrino masses, matter-antimatter asymmetry and dark matter.}

\FullConference{The 19th International Workshop on Neutrinos from Accelerators-NUFACT2017\\
		25-30 September, 2017\\
		Uppsala University, Uppsala, Sweden}

\begin{document}

\section{Introduction}

The LHC experiments have been running for several years without finding new physics at the TeV scale. A complementary approach is to probe the presence of new particles at lower energy scales with couplings to the Standard Model so weak that they have escaped detection in previous searches. In models with hidden sectors, there are many well-motivated candidates which could be produced at high-intensity beams through various portals~\cite{Alekhin2015}, such as axions, dark photons, dark scalars, and right-handed neutrinos (hereafter termed heavy neutral leptons, or HNLs). 

\begin{figure}[htb]
\centering
\includegraphics[width=0.49\linewidth]{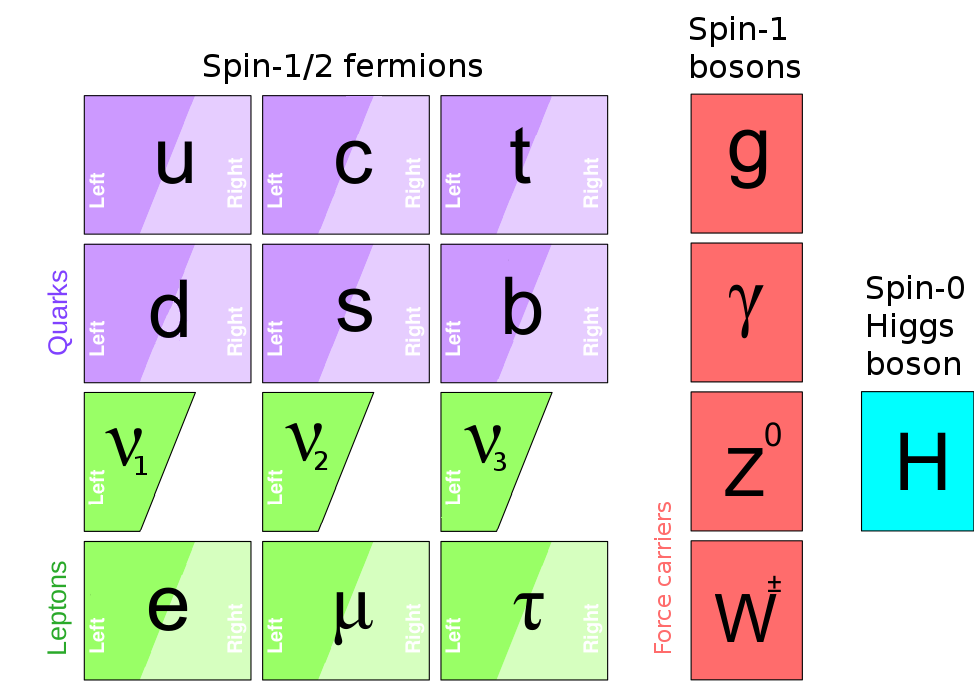}
\includegraphics[width=0.49\linewidth]{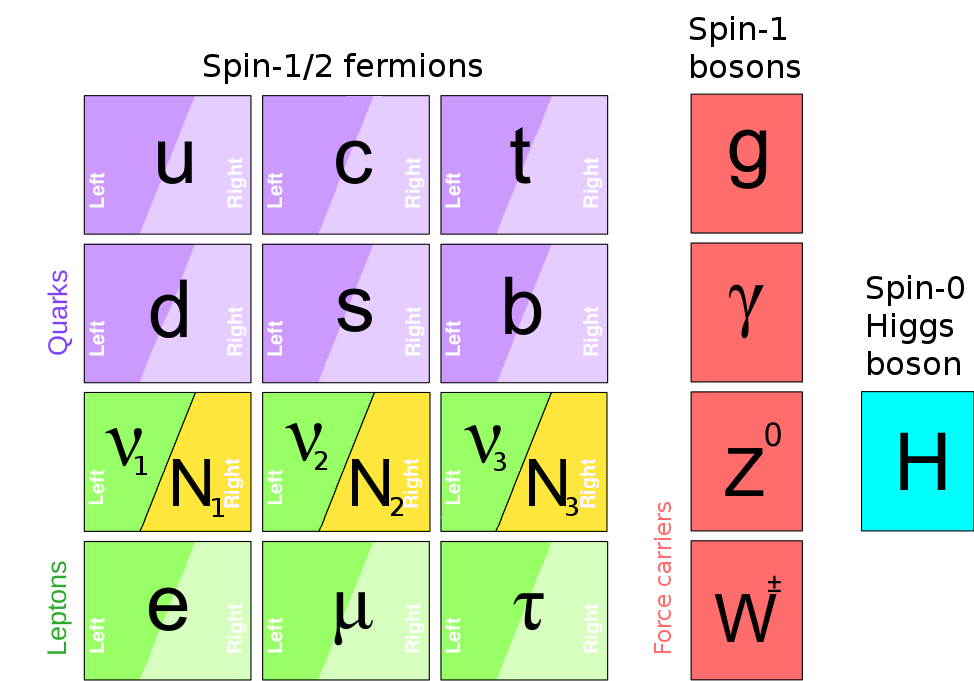}
\caption{Left: summary of the particle states in the Standard Model, indicating left-handed and right-handed fermions separately. All particles in this table have been experimentally observed. Right: three right-handed neutrinos $N_{1,2,3}$ are added and given Majorana masses below the electroweak scale to solve the problems of neutrino masses, matter-antimatter asymmetry in the Universe, and dark matter~\cite{Asaka2005b}. }
\label{fig:nuMSM}
\end{figure}

Remarkably, the hypothesis of three HNLs with Majorana masses below the electroweak scale in combination with CP violation in the neutrino sector can at once explain neutrino masses through the seesaw mechanism, explain the excess of matter over antimatter in the Universe through leptogenesis, and provide a valid candidate for dark matter~\cite{Asaka2005b,Canetti2013a} (see Fig.~\ref{fig:nuMSM}). HNLs would interact only via gravitation and a tiny mixing to neutrinos. This means that HNL production in a laboratory would only be possible at the highest beam intensities~\cite{Gorbunov2007} and HNLs would typically have long lifetimes, leading to a signature of a displaced decay.

The high-intensity 400~GeV proton beam of the CERN SPS provides a suitable facility for producing high yields of hadrons which could decay into hidden particles. To uncover these, the identification of decay vertices in experiments such as NA62 and SHiP provides powerful search strategies. 

\section{NA62 in beam mode} 
\label{NA62beam}

\begin{figure}[htb]
\centering
\includegraphics[width=0.99\linewidth]{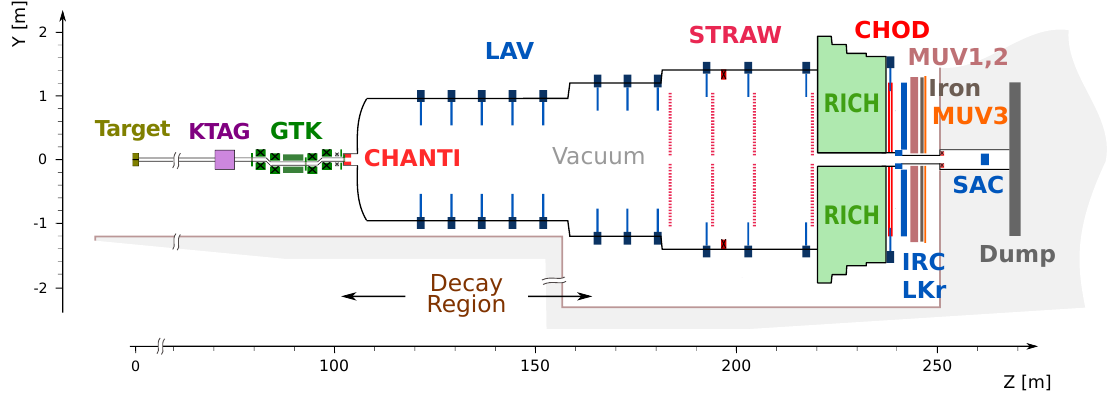}
\caption{Outline of the NA62 experiment in beam mode~\cite{NA622017a}, for reconstructing kaon decays.}
\label{fig:Na62}
\end{figure}

In usual operation, the NA62 experiment~\cite{NA622017a} makes use of the SPS proton beam on a target to produce a collimated mixed beam of positively charged pions ($\pi^+$), kaons ($K^+$) and protons ($p$) of 75~GeV momentum. A kaon tagger detector (KTAG) is used to identify the presence of individual $K^+$ particles inside the beam. A vacuum vessel starting around 100~m downstream of the target serves as a decay volume and contains the main tracking system made of straw tubes that operate in vacuum, as illustrated in Fig.~\ref{fig:Na62}. Further detectors allow precise reconstruction and identification of the $K^+$ decay products.

\begin{figure}[htb]
\centering
\includegraphics[width=0.3\linewidth]{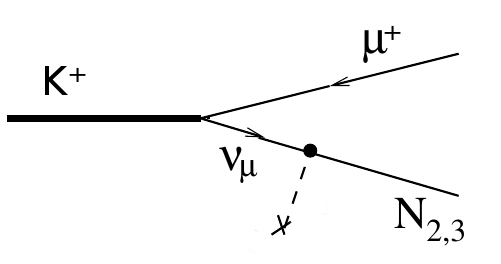}
\caption{Kaon decay into a muon and an HNL. In the NA62 search in beam mode, the HNL escapes detection and its mass is reconstructed using the muon and kaon kinematics.}
\label{fig:Kdecay}
\end{figure}

Leptonic $K^+$ decays produce a neutrino which can potentially mix with an HNL with mass up to 0.45~GeV, as illustrated in Fig.~\ref{fig:Kdecay}. NA62 is capable of measuring the kinematics of the charged lepton from the $K^+$ decay with high precision. The KTAG detector ensures that the mother particle is a 75~GeV $K^+$ to further constrain the two-body decay kinematics, allowing to reconstruct the mass of the neutrino. Using this technique, searches were made using $6\cdot 10^7$ and $3\cdot 10^8$ $K^+$ decays in the muon and electron channels, respectively. No excesses over backgrounds were seen in the neutrino mass distributions, allowing to set limits in the coupling strength $U^2$ (mixing with $\nu_\mu$) in the range $10^{-5}-10^{-6}$ in the mass range $0.3-4.5$~GeV~\cite{NA622017b}, and preliminary limits in $U^2$ (mixing with $\nu_e$) the range $10^{-6}-10^{-7}$ in the mass range $0.16-4.5$~GeV. These limits are competitive when considering searches in which no assumptions are made about the HNL decays, while searches in dump mode in which HNL decays are identified explicitly provide stronger constraints in this mass range~\cite{Bernardi1988}.

Hadronic $K^+$ decays with a $\pi^0$ in the final state such as $K^+\rightarrow \pi^+\pi^0$ can be used to search for dark photons. In this case, one of the $\gamma$s from the $\pi^0$ decay is assumed to mix with a dark photon and the dark photon mass can be reconstructed from the kinematics of the $K^+$, $\pi^+$ and $\gamma$. Preliminary results using 5\% of the 2016 dataset constrain unexplored regions of the parameter space for dark photon masses around 0.1~GeV~\cite{Lanfranchi2017}. The sensitivity will further improve as more data are analysed.   

\section{NA62 in dump mode}
\label{NA62dump}

\begin{figure}[htb]
\centering
\includegraphics[width=0.7\linewidth]{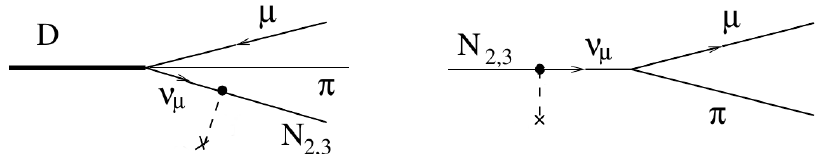}
\caption{Typical processes by which an HNL is produced through charmed hadron decay (left) and decays back into charged particles after some distance (right). The decay can be reconstructed as a displaced vertex in NA62 in off-target mode or in SHiP.}
\label{fig:Ddecay}
\end{figure}

A promising strategy to search for HNLs with masses of the order of the GeV is production through hadron decays at high-intensity proton beam-dump facilities. Searches of this type can access HNL masses up to 2~GeV using charmed hadron decays with reconstruction of the HNL decay at some distance from the interaction point (see Fig.~\ref{fig:Ddecay}). The best constraints to date in the mass range $0.45-2$~GeV were obtained using this technique with the CHARM experiment at CERN~\cite{CHARM1986} and the NuTeV experiment at Fermilab~\cite{NuTeV1999}.  

NA62 can be operated in so-called dump mode by lifting the target and letting the beam impinge directly on the Cu collimator placed 20~m downstream of the target. This operation does not disrupt the setup, thus allowing to switch easily back to beam mode. A test run was performed in dump mode with $2\cdot 10^{15}$ protons on target to study backgrounds which could fake vertices in an HNL search. In this test, HNL candidates were selected by using vertices made of opposite-charge tracks within a 1~ns window, and the reconstructed points of origin of the HNL candidates were found to be at a distance larger than 5~cm from the interaction point, thus failing to enter the signal region. The backgrounds are dominated by muons produced upstream of the vessel and can be further rejected by adding an upstream veto detector to the NA62 setup. 

\begin{figure}[t]
\centering
\includegraphics[width=0.98\linewidth]{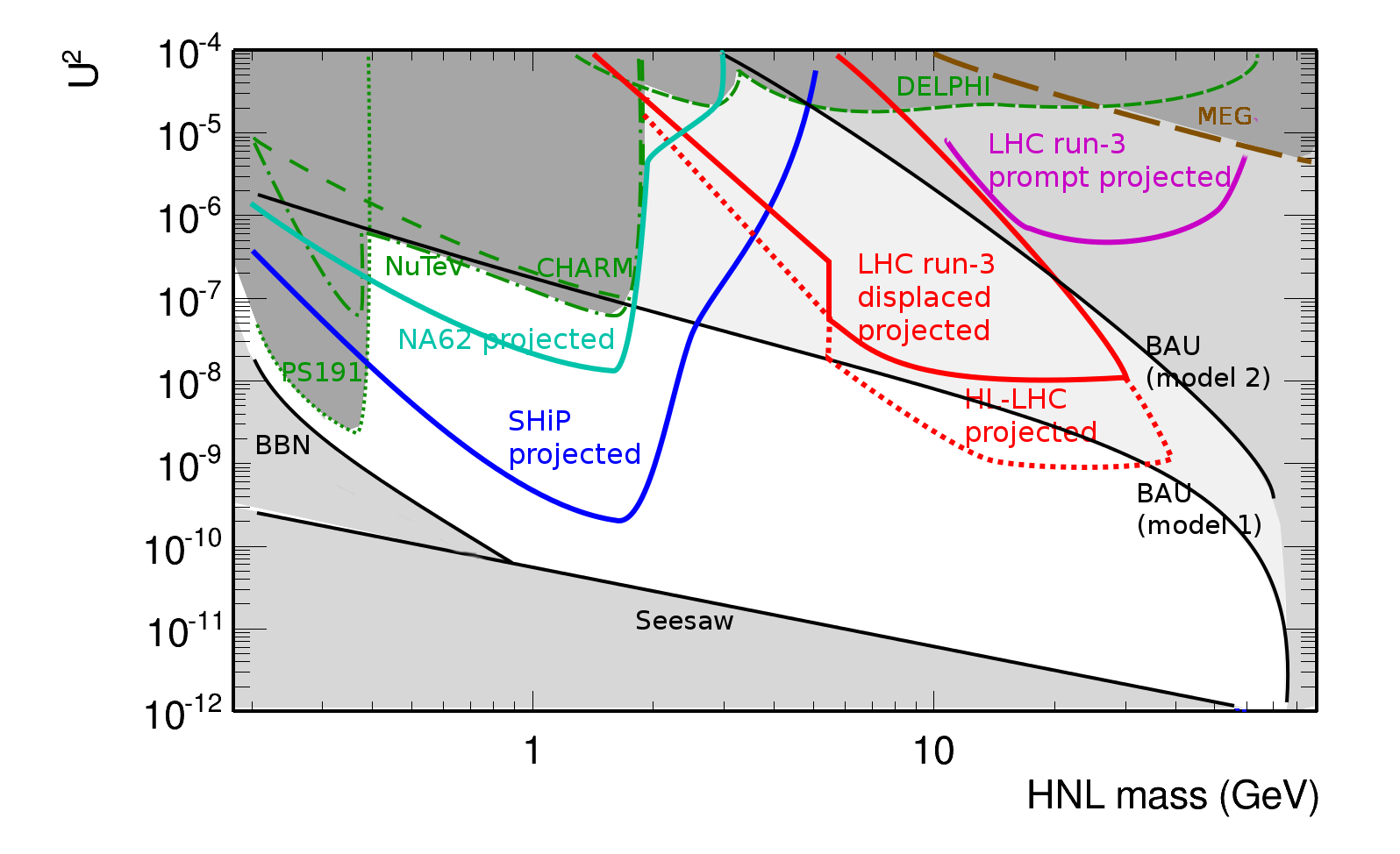}
\caption{Estimates of experimental sensitivities to HNLs at NA62 in dump mode and SHiP, in the coupling strength ($U^2$ for mixing to $\nu_\mu$) vs. mass plane. Direct~\cite{Bernardi1988,CHARM1986,NuTeV1999,Delphi1997,CMS2015b} and indirect~\cite{MEG2013,Antusch2015} experimental limits are indicated as dashed lines. Details about the LHC projections are given in Ref.~\cite{Mermod2017}.}
\label{fig:sensitivity}
\end{figure}

These tests provide confidence that a dedicated HNL search at NA62 in dump mode can access regions of the parameter space beyond the limits sets by the CHARM and NuTeV experiments, as shown in Fig.~\ref{fig:sensitivity} -- here assuming dominant HNL mixing to muon neutrinos, detection of all two-track final states with consideration of geometrical acceptance and trigger efficiency, zero backgrounds, and a dataset corresponding to $10^{18}$ protons on target~\cite{Lanfranchi2017}. Likewise, NA62 in dump mode can also probe dark photons in regions of the parameter space beyond previous constraints. The aim is to accumulate $10^{18}$ protons on target during 3 months of dedicated data taking in $2021-2023$. 

\section{The SHiP experiment}
\label{SHiP}

SHiP is a general-purpose fixed-target facility proposed at the CERN SPS to search for particles with very low couplings to the Standard Model~\cite{Bonivento2013,CERN2014,SHiP2015}. The 400~GeV proton beam extracted from the SPS will be dumped on a high density target with the aim of accumulating $2\times 10^{20}$ protons on target during 5 years of operation. It will produce a large number of neutrinos through hadron decays, following the same principle as described in Section~\ref{NA62dump}. In particular, neutrinos from decays of hadrons containing $c$ or $b$ quarks can potentially mix with HNLs with masses up to 5~GeV. The charm production at SHiP, with an expected total of $\sim 5\cdot 10^{16}$ neutrinos produced in charm decays, largely surpasses that of any other existing or planned facility, allowing to probe very small coupling strengths and resulting in the HNLs, if produced, to travel very large distances until they decay. 

\begin{figure}[htb]
\centering
\includegraphics[width=0.99\linewidth]{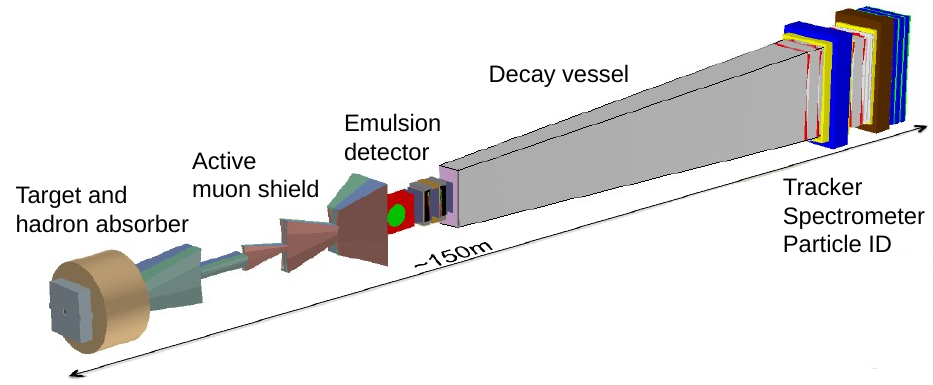}
\caption{Current design of the SHiP experiment.}
\label{fig:SHiP}
\end{figure}

In its current design, SHiP comprises a target followed by a hadron absorber, a muon shield, a 50~m long, 5$-$10~m wide decay volume and a set of detectors for track reconstruction and particle identification, as shown in Fig.~\ref{fig:SHiP}. The active muon shield is a set of magnets designed to minimise the flux of muons entering the vessel while allowing to have the vessel as close as possible to the target~\cite{SHiP2017a}. The experiment as a whole is optimised to reconstruct and identify decays from new long-lived neutral particles and reject backgrounds which could mimic such decays~\cite{SHiP2015,Alekhin2015}. 

To reduce backgrounds in SHiP, neutrino production in the forward direction is reduced by stopping hadrons in a dense absorber before they decay, neutrino interactions are minimised by evacuating the air in the vessel, and charged particles entering the vessel are vetoed by surrounding the decay volume with tagging detectors. One additional handle to reject random crossings is to measure and match the arrival times of the particles forming the vertex with a high precision ($100$~ps resolution or better) using a dedicated timing detector, typically made of bulk scintillator bars~\cite{Betancourt2017}. Simulations show that the combined use of the active muon shield, veto taggers surrounding the vessel, timing detector, track momentum and pointing measurements, and muon-pion separation, can reduce the backgrounds to 0.1 events in a sample of $2\times 10^{20}$ protons on target~\cite{SHiP2015}. 

A preliminary estimate of the SHiP sensitivity to HNLs is shown in Fig.~\ref{fig:sensitivity}. SHiP is a state-of-the-art experiment for hidden sector searches and will clearly be able to dig deeply into the most favoured regions of the HNL parameter space. The SHiP sensitivity for dark photons, as well as many other long-lived particles in scenarios of new physics with hidden sectors, also goes well beyond what could be previously explored.  

SHiP will also be equipped with an emulsion cloud chamber detector followed by a muon spectrometer, placed upstream of the decay volume. This will allow to study the interactions of tau neutrinos with unprecedented precision and also the direct search for interactions of dark-matter particles produced in the beam dump. 

\section{Summary and outlook}

The NA62 and SHiP experiments at the CERN SPS are capable of probing new particles with very small couplings to the Standard Model with unprecedented sensitivity. NA62 in beam mode already set competitive limits for HNLs and dark photons, with a sensitivity which will be improved in the near future as more data is being analysed. Tests made with NA62 in dump mode provide insights about the expected backgrounds in searches for hidden particle decays. Runs to be scheduled at NA62 in dump mode during the next few years provide very promising prospects in particular for HNLs and dark photon searches. The SHiP experiment, with its very high fluxes, large decay volume and powerful background rejection systems, is a supreme experiment to search for hidden particles at a proton beam-dump facility. SHiP will dig deep into new territories in a plethora of new physics scenarios.

\bibliographystyle{mystylem}
\bibliography{HNLrefs}


\end{document}